\documentclass[aps,pra,twocolumn,superscriptaddress,longbibliography]{revtex4-2}

\pdfoutput=1
\usepackage[english]{babel}
\usepackage{color}
\newcommand{\comment}[1]{}
\usepackage{amsmath}
\usepackage{amsfonts}
\usepackage{tensor}
\usepackage{physics}
\usepackage{graphicx}
\usepackage[colorlinks=true, allcolors=blue]{hyperref}
\usepackage{siunitx}

\begin{document}

\title{Quantum Gate Optimization for Rydberg Architectures in the Weak-Coupling Limit}
\author{Nicolas Heimann}
\email{nheimann@physnet.uni-hamburg.de}
\affiliation{Zentrum f\"ur Optische Quantentechnologien, Universit\"at Hamburg, 22761 Hamburg, Germany}
\affiliation{Institut für Quantenphysik, Universit\"at Hamburg, 22761 Hamburg, Germany}
\affiliation{The Hamburg Centre for Ultrafast Imaging, 22761 Hamburg, Germany}
\author{Lukas Broers}
\affiliation{Zentrum f\"ur Optische Quantentechnologien, Universit\"at Hamburg, 22761 Hamburg, Germany}
\affiliation{Institut für Quantenphysik, Universit\"at Hamburg, 22761 Hamburg, Germany}
\author{Nejira Pintul}
\author{Tobias Petersen}
\author{Koen Sponselee}
\affiliation{Zentrum f\"ur Optische Quantentechnologien, Universit\"at Hamburg, 22761 Hamburg, Germany}
\affiliation{Institut für Quantenphysik, Universit\"at Hamburg, 22761 Hamburg, Germany}
\author{Alexander Ilin}
\affiliation{Zentrum f\"ur Optische Quantentechnologien, Universit\"at Hamburg, 22761 Hamburg, Germany}
\affiliation{Institut für Quantenphysik, Universit\"at Hamburg, 22761 Hamburg, Germany}
\affiliation{The Hamburg Centre for Ultrafast Imaging, 22761 Hamburg, Germany}
\author{Christoph Becker}
\affiliation{Zentrum f\"ur Optische Quantentechnologien, Universit\"at Hamburg, 22761 Hamburg, Germany}
\affiliation{Institut für Quantenphysik, Universit\"at Hamburg, 22761 Hamburg, Germany}
\author{Ludwig Mathey}
\affiliation{Zentrum f\"ur Optische Quantentechnologien, Universit\"at Hamburg, 22761 Hamburg, Germany}
\affiliation{Institut für Quantenphysik, Universit\"at Hamburg, 22761 Hamburg, Germany}
\affiliation{The Hamburg Centre for Ultrafast Imaging, 22761 Hamburg, Germany}

\begin{abstract}
We demonstrate machine learning assisted design of a two-qubit gate in a Rydberg tweezer system. Two low-energy hyperfine states in each of the atoms represent the logical qubit and a Rydberg state acts as an auxiliary state to induce qubit interaction. Utilizing a hybrid quantum-classical optimizer, we generate optimal pulse sequences that implement a $\mathrm{CNOT}$ gate with high fidelity, for experimentally realistic parameters and protocols, as well as realistic limitations.
We show that local control of single qubit operations is sufficient for performing quantum computation on a large array of atoms.
We generate optimized strategies that are robust for both the strong-coupling, blockade regime of the Rydberg states, but also for the weak-coupling limit. Thus, we show that Rydberg-based quantum information processing in the weak-coupling limit is a desirable approach, being robust and optimal, with current technology.
\end{abstract}
\maketitle

\begin{figure*}
    \centering
    \includegraphics[width=1.0\linewidth]{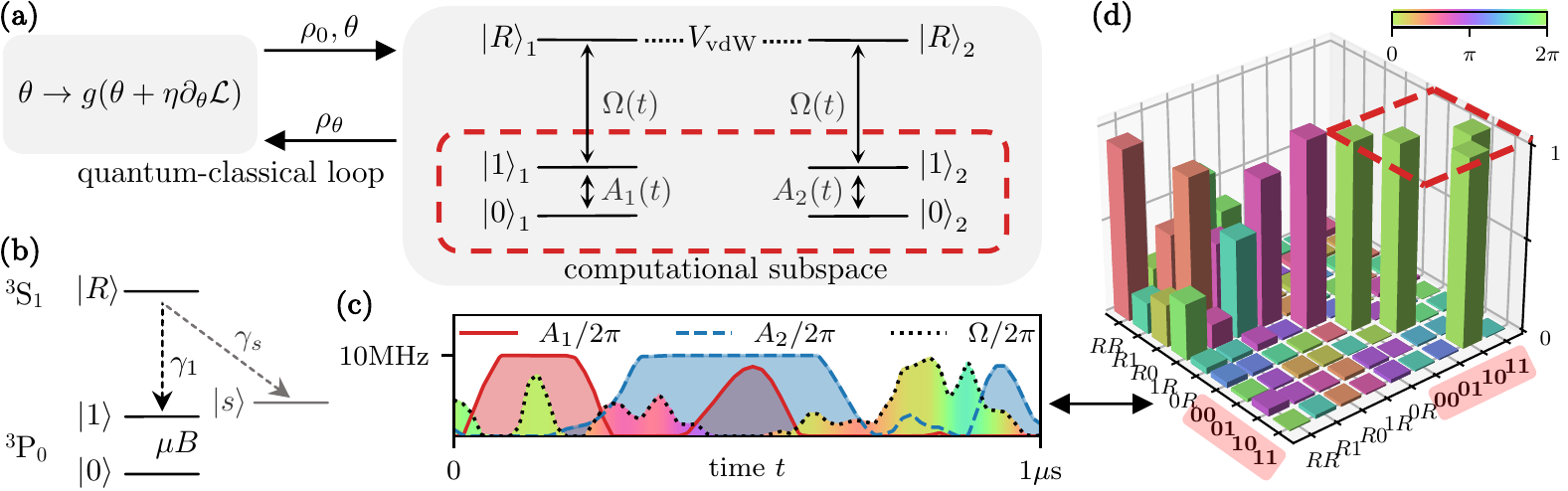}
    \caption{
    \textbf{Optimization platform.}
    \textbf{(a)} Hybrid quantum-classical optimization scheme. The hyperfine states $\ket{0}$ and $\ket{1}$ act as logical states, forming the computational subspace depicted as red-dashed lines, and can be manipulated individually by the control Raman protocol $A_1(t)$ and the target Raman protocol $A_2(t)$, in which qubit 1 is the control qubit and qubit 2 is the target qubit. The Rabi protocol $\Omega(t)$ controls transfer between the state $\ket{1}$ and a highly excited Rydberg state $\ket{R}$.
    The protocols $\Omega(t)$, $A_j(t)$ and $B$ are parameterized by the transformation parameters $\theta=\{\theta_i\}$.
    The atoms are coupled via the van-der-Waals interaction $V_\text{vdW}$ of the Rydberg states. For a transformation parameter set $\theta$, an initial state $\rho_0$ is propagated yielding the final state $\rho_\theta$. This time propagation in the quantum unit is controlled by a classical unit synthesising the loss function $\mathcal{L}$ from $\rho_\theta$, allowing to optimize the transformation parameters $\theta$.
    \textbf{(b)} Level diagram and dissipation channels, for our main example of $^{171}\textrm{Yb}$. We consider decay from $\ket{R}$ into $\ket{1}$, as well as into $\ket{s}$ which is otherwise decoupled dynamically, and is introduced to model population loss. The magnetic field $B$ defines the Zeeman splitting of the two hyperfine states.
    \textbf{(c)}~A high fidelity protocol implementing the $\mathrm{CNOT}$ transformation. The Rabi protocol $\Omega(t)$ and the Raman protocols $A_j(t)$ are constrained by a maximal frequency of $\Omega_\text{max}=A_\text{max}=2\pi\times 10\si{\mega\hertz}$.
    The phase $\phi(t)$ of the Rabi protocol $\Omega(t)$ is depicted by the filling color.
    \textbf{(d)}~The transformation $U$ corresponding to the pulse sequence shown in (c), depicted at the time $\tau=1\si{\micro\second}$. The $\mathrm{CNOT}$ operation is clearly visible in the computational subspace, enclosed by the red-dashed square.}
    \label{fig:scheme}
\end{figure*}
\section{Introduction}
Rydberg tweezer arrays have evolved into an intriguing and promising platform for quantum computing~\cite{saffman_quantum_2016, cohen_quantum_2021, graham_multi-qubit_2022} and quantum simulation~\cite{whitlock_quantum_2021, browaeys_many-body_2020}.
These devices support the preparation of scalable, nearly defect-free systems~\cite{hundreds-atoms-22, berredo_atombyatom_2016, endres_atom-by-atom_2016}, high fidelity single-qubit operations~\cite{madjarov_high-fidelity_2020} and the implementation of two-qubit gates via Rydberg states~\cite{jaksch_fast_2000, lukin_dipole_2001, saffman_analysis_2005, urban_observation_2009, wilk_entanglement_2010}.
This includes the quantum gate design based on Rydberg blockade, that corresponds to the widely explored regime of strong van-der-Waals interaction strength and small interatomic distances. Furthermore, qubit architectures based on alkaline-earth and alkaline-earth-like atoms~\cite{chen_analyzing_2022}, such as strontium~\cite{cooper_alkaline-earth_2018, norcia_microscopic_2018} and ytterbium atoms~\cite{wilson_trapped_2022, jenkins_ytterbium_2022, ma_universal_2022}, have desirable features such as long-lived decoupled nuclear spin states that are suitable to be used as qubit states, as well as single-photon Rydberg transitions for implementing fast two-qubit gates. The existence of a meta-stable clock state further allows for elaborate qubit schemes allowing novel error correction strategies and shelving operations for non-destructive mid-circuit readout~\cite{wu_erasure_2022}.
Further design options include triple magic trapping of qubit and Rydberg states~\cite{triple-magic-16}, and local Rydberg control via manipulation of inner shell electrons~\cite{local-control-22}.

Optimization methods, such as quantum machine learning and quantum optimal control are a powerful and versatile approach of operating and controlling quantum dynamics in a way that is optimal or near-optimal according to a desired metric. In particular, variational quantum algorithms~\cite{cerezo_variational_2021, choquette_quantum_optimal_2021, yang_optimizing_2017} are a class of algorithms which utilize a generalized quantum circuit with parameterized gates to transform the synthesis of quantum algorithm solutions into an optimization problem.
This approach can be extended towards quantum optimal control~\cite{Brif_2010, caneva_chopped_2011, magann_from_pulses_2021, koch_quantum_2022, lbroers_barren_2022} and has been utilized in different noisy intermediate-scale quantum devices~\cite{Preskill2018quantumcomputingin}, such as trapped ions \cite{kang_batch_optimization_2021, figgatt_parallel_2019, choi_optimal_2014, nebendahl_optimal_2009}, superconducting qubits \cite{werninghaus_leakage_2021, huang_optimal_2014, egger_2013_optimizing, rebentrost_optimal_2009} and neutral atoms \cite{jandura_time-optimal_2022, mohan_robust_2022, jandura_optimizing_2022, goerz_robustness_2014, mueller_optimizing_2011}.
Recently, time-optimal gates have been constructed using quantum optimal control~\cite{jandura_time-optimal_2022} and realized experimentally~\cite{evered2023highfidelity}.

In this paper, we demonstrate machine learning assisted design of a controlled-not (CNOT) gate in Rydberg tweezer systems. The logical qubit states are implemented in two hyperfine states of the atoms, which are controlled via Raman pulses. Additionally we consider a Rydberg state in each atom, which can be Rabi-driven from one of the hyperfine states.
We demonstrate that using either a global Rabi protocol, driving the Rydberg transition of all atoms, and individual Raman protocols, driving the hyperfine transition of individual atoms, or a global Raman protocol and individual Rabi protocols, are sufficient to support universal quantum computing.
We focus on the case of a global Rabi protocol and individual Raman protocols. The parameters of the atomic states and the magnitudes of the Rabi and Raman protocols, as well as an applied magnetic field, are modeled after $^{171}\textrm{Yb}$ atom tweezers. However, we emphasize that our analysis and results are directly applicable to all Rydberg tweezer systems, as they include realistic conditions of operation of current devices. We consider a fixed total operation time, and determine fidelity-optimal implementations based on a hybrid quantum-classical optimizer algorithm.
We use the van-der-Waals interaction strength as a variable parameter. We identify the minimal van-der-Waals interaction that supports an implementation of a CNOT gate with high fidelity, and find that the fidelity saturates beyond that magnitude.
We determine the robustness of our optimal implementations with respect to fluctuations of the distance between the atoms.
We find that the implementations are not only robust in the blockade but also in the weak-coupling limit.
We propose this regime to be utilized for robust optimal quantum computing under realistic conditions with current technology.

This paper is organized as follows.
In Sect. II, we introduce the model and method used throughout the manuscript.
In Sect. III, we present the performance and the protocols of the hybrid quantum-classical optimizer within the weak-coupling limit.
In Sect. IV, we show how spatial fluctuations affect the gate fidelity for a realistic range of interatomic distances. 
In Sect. V, we conclude.


\section{Model}
We consider neutral atoms trapped individually in optical tweezers. For each of the atoms we consider two long-lived, low-energy states that constitute a qubit, written as $\ket{0}$ and $\ket{1}$. Additionally we consider a highly excited Rydberg state $\ket{R}$, and a generic state $\ket{s}$ that we use to model the decay of the Rydberg state. The Rydberg state $\ket{R}$ is utilized for its strong van-der-Waals interaction between two atoms in this state, providing a non-linearity to design two-qubit gates. We consider the Hamiltonian
\begin{equation}
    H = \sum_j H_j + \sum_{i,j}V^{i,j}_{\mathrm{vdW}} \ket{R}_i\ket{R}_j\bra{R}_i\bra{R}_j,
    \label{eq:hamilton}
\end{equation}
where 
\begin{equation}
V^{i,j}_\mathrm{vdW} = \frac{\hbar C_6}{|r_i-r_j|^6}
\label{eq:vdw}
\end{equation}
is the van-der-Waals interaction between the $i$th and $j$th atom at the respective positions $r_i$ and $r_j$. $C_6$ is the coefficient of the van-der-Waals interaction and depends on the specific atom species and Rydberg state. 
We choose $C_6=1 \si{THz} \cdot\si{\micro\metre}^{6}$, as inspired by~\cite{chen_analyzing_2022, ma_universal_2022}, as a typical magnitude for Rydberg atoms of two-electron atoms. $H_j$ is the local Hamiltonian of the $j$th atom. It is
\begin{equation}
    H_j = \frac{\hbar}{2}\begin{pmatrix}
        0 & \Omega(t) & 0 \\ \Omega^*(t) & 0 &  A_j(t) \\ 0 &  A_j(t) & 0
    \end{pmatrix}
    +\frac{1}{2}\begin{pmatrix}
        0 &0 & 0 \\ 0 & \mu B & 0 \\ 0 & 0 & -\mu B
    \end{pmatrix},
\end{equation}
and operates on the states $\{\ket{R},\ket{1}, \ket{0}\}$.
The two logical qubit states $\ket{0}$ and $\ket{1}$ are realized as hyperfine states of the atom, and $\mu B$ is the Zeeman splitting between them induced by an external magnetic field $B$, where $2\mu$ is the difference of the magnetic moments. The corresponding Zeeman shift of the Rydberg state is normalized to zero in the rotating frame.
$A_j(t)$ is the Raman coupling between the logical qubit states $\ket{0}_j$ and $\ket{1}_j$ of atom $j$, which derives from a two-photon transition, and which we consider to be real-valued. This assumption is realized by optimizing the excitation light homogeneity to select the phases of each coupling $A_j(t)$~\cite{endres_atom-by-atom_2016} to zero. $\Omega(t) = |\Omega(t)|e^{-i\phi(t)}$ is the complex-valued Rabi coupling between the levels $\ket{1}_j$ and $\ket{R}_j$ for all $j$, i.e. in a global fashion. 
As discussed in App.~\ref{appendixa}, we show that using either a global Rabi coupling and individual Raman couplings, or indvidual Rabi couplings and a global Raman coupling, is sufficient for universal quantum computing.
We illustrate the hybrid quantum-classical optimizer in Fig.~\ref{fig:scheme}.
For a given optimization task, we limit the Rabi and Raman coupling by maximal values of $\Omega_\textrm{max}$ and $A_\textrm{max}$. Here, we focus on the case of $\Omega_\textrm{max}=A_\textrm{max}=2\pi\times 10\si{MHz}$, but note that our approach equally applies to the general case, i.e. $\Omega_\textrm{max}\neq A_\textrm{max}$.
In App.~\ref{appendix:omega-max}, we show that a value of $\Omega_\textrm{max}=2\pi\times 10\si{MHz}$ is sufficient for our analysis.
The maximum gradient of the Rabi phase $\phi(t)$ is $\partial_t\phi_\textrm{max}=\pm\pi/100\si{\nano\second}$, motivated by typical acousto-optic modulator bandwidths.
The magnetic field $B$ is assumed to be stationary, but can be chosen arbitrarily within the range $B_\textrm{min} \leq  B \leq B_\textrm{max}$, where $B_\textrm{min}=100\mathrm{G}$ and $B_\textrm{max}=200\mathrm{G}$ which corresponds to $0.1\si{\mega\hertz} \leq \mu B/h \leq 0.2\si{\mega\hertz}$.

The finite lifetime $1/\gamma$ of the Rydberg state leads to decoherence.
We consider two contributions to the decay, black body radiation and spontaneous decay~\cite{cong_hardware-efficient_2022}.
Both black body radiation and spontaneous decay induce transitions out of the Rydberg state $\ket{R}$ to states other than $\ket{0}$, $\ket{1}$, and $\ket{R}$. We model these states with an auxiliary state $\ket{s}$.
Spontaneous decay also induces transitions from the Rydberg state $\ket{R}$ to the state $\ket{1}$, i.e. the computational subspace.
In Fig.~\ref{fig:scheme}~(b) we illustrate this effective dissipation model.
The dynamics of the system are governed by the Lindblad master equation
\begin{equation}
    \dot{\rho} = -\frac{i}{\hbar}[H,\rho] 
        + \sum_{ij} \mathcal{D}[L_{i}^{j}]\rho,
\label{eq:drho}
\end{equation}
where $\mathcal{D}[L]\rho=L\rho L^\dagger-\frac{1}{2}\{L^{\dagger} L, \rho\}$, with the Lindblad operators $L_j^s~=~\sqrt{\gamma_s}\ket{s}_j\bra{R}_j$ and $L_j^1~=~\sqrt{\gamma_1}\ket{1}_j\bra{R}_j$ of the $j$th atom.
The total decay rate of the Rydberg state obeys $\gamma=\gamma_s+\gamma_1$ where $\gamma_s=20 \gamma_1$~\cite{wu_erasure_2022}. Here we choose typical values of the lifetime of the Rydberg state of $1/\gamma=10\si{\mu s}$, $100\si{\mu s}$ and $500\si{\mu s}$ ~\cite{chen_analyzing_2022}.
We assume magic-trapping between $\ket{1}$ and $\ket{R}$ and neglect losses arising from turning off the trap during gate operations as well as dephasing contributions~\cite{wilson_trapped_2022, zhang_magic_2011}.

Gradient Ascent Pulse Engineering (GRAPE)~\cite{khaneja_optimal_2005} is a quantum optimal control technique to construct pulse sequences, which determine the dynamical evolution of the system, such that a desired target transformation $U$ is realized.
Note that we employ this method for non-unitary dynamics, given in Eq.~\ref{eq:drho}.
We consider a general Hamiltonian $H_\theta(t) = H_0 + \sum_{k} f_k(t; \theta) h_k + \text{h.c.}$, where $f_k(t; \theta)$ are complex-valued functions, $\theta=\{\theta_i\}$ are parameters, which are to be optimized, and the $h_k$ are hermitian operators.
The transformation parameters $\theta$ correspond to a transformation which we evaluate for a given state $\rho_0$ by integrating Eq.~(\ref{eq:drho}) over the algorithm time $\tau$.
We denote the final state a particular parameter set $\theta$ as $\rho_\theta$.
Throughout this work, we fix the algorithm time to $\tau=1\si{\micro\second}$.
The optimization is performed with respect to the objective, i.e. the loss function, which in our case we define as
\begin{equation}
    \mathcal{L} = 1-F_\theta = 1-\frac{1}{4} \langle|\text{Tr}(\rho_\theta^\dagger PU\rho_0 U^\dagger P)|\rangle_{\rho_0},
 \label{eq:loss}
\end{equation}
where $F_\theta$ is the fidelity and $P=\sum_q \ket{q}\bra{q}$ is the projector onto the computational subspace, which is $P=\ket{00}\bra{00}+\ket{01}\bra{01}+\ket{10}\bra{10}+\ket{11}\bra{11}$.
$\langle\cdot\rangle_{\rho_0}$ is the average over $32$ initial random product states $\rho_0=\bigotimes_i \rho_i$ sampled from the Bloch spheres of the computational subspaces.
The batch size of 32 is an empirical value that provides efficient optimization.
The optimal transformation parameters $\theta_\textrm{opt}=\mathrm{argmin}_{\theta} \mathcal{L}$ are inferred via stochastic gradient descent~\cite{Goodfellow-2016}.
First, the loss $\mathcal{L}$ is evaluated given the transformation parameters $\theta$.
Next, the parameters are varied as $\theta_i\rightarrow\theta_i+\epsilon$ by a small amount $\epsilon=10^{-8}$, and subsequently the modified loss $\mathcal{L}_{i}$ is evaluated.
The first order gradient is approximated by the finite difference $\partial \mathcal{L}/\partial \theta_i=(\mathcal{L}_i-\mathcal{L})/\epsilon$ and the parameters are then updated as
\begin{equation}
    \theta_i \rightarrow g_i\left(\theta_i + \eta_i\frac{\partial\mathcal{L}}{\partial \theta_i}\right),
    \label{eq:param-step}
\end{equation}
where $\eta_i$ are dynamically adapted learning rates according to the ADAM method~\cite{kingma_adam_2017}.
The functions $g_i$ impose constraints on the protocols.
Note that these constraints do not affect the gradient.
We refer to this step in the algorithm as a training epoch and illustrate this in Fig.~\ref{fig:scheme}~(a). Optimization occurs by iterating over the training epochs until convergence.

The central example that we apply this optimization method to, is the optimal implementation of the CNOT gate. So the number of atoms $N_a = 2$. 
However, we emphasize that the methodology presented here naturally applies to atom systems with larger numbers.
The notion of a global Rabi coupling implies that for $N_a>2$, any additional atom besides the two involved in the $\mathrm{CNOT}$ operation will also experience the global coupling $\Omega(t)$. This results in a transformation on these other qubits, which may be undesired.
Our optimization method can also be utilized to learn a coupling $A_{j>2}(t)$ that implements the identity operation, in the presence of the fixed global coupling $\Omega(t)$.
We emphasize that this is possible, because even in the case of arbitrarily many neutral atoms, a single global Rabi coupling is sufficient for universal quantum computing. 
For example, the resulting transformation on the other qubits can be mitigated efficiently by moving the other atoms sufficiently far apart such that the van-der-Waals interaction becomes negligible while additionally applying the control coupling $A_1(t)$. By construction of the $\mathrm{CNOT}$ gate, the control coupling $A_1(t)$ will transform the qubit states $\ket{0}$ and $\ket{1}$ into themselves, respectively. This will in general only result in a relative phase between these states which can be corrected.
Alternatively, for alkaline-earth-like atoms like $^171$Yb, the omg qubit architecture~\cite{chen_analyzing_2022} can be employed to realize local two-qubit gates despite the global coupling $\Omega(t)$. Since the Rydberg excitation originates from the meta-stable $^3P_0$ state it is straight forward to site-selectively shelve atoms in the ground state qubit $^1S_0$ if the CNOT gate is not desired.
We note that universal quantum computing is equally possible in the case in which there is a global Raman coupling that is equal for all atoms, and the Rabi couplings are applied to the atoms individually. This result is both conceptually interesting, as well as of experimental relevance, because it suggests an alternative, minimal set of experimental control parameters. We expand on this implementation approach elsewhere. In this work, we focus on the case of a global Rabi pulse. For details on the computational universality under these constraints, see App.~\ref{appendixa}.

We optimize the transformation parameters $\theta$, which parameterize $\Omega(t)$ and $A_j(t)$ as stepwise functions which we linearly interpolate in the dynamics, as well as $B$ which we consider to be constant during the time evolution such that it is represented by a single parameter. We refer to these parameterizations as the Rabi protocol $\Omega(t)$, the control Raman protocol $A_1(t)$ and the target Raman protocol $A_2(t)$.
We give a detailed account of the parameterization in App.~\ref{appendix:protocol-parameterization}.
We construct the initial protocols $|\Omega^0(t)|$ and $A_j^0(t)$ to be positive and slowly varying. The initial phase of the Rabi protocol $\phi^0(t)$ is generated via a random walk starting at $\phi^0(0)=0$, see App.~\ref{appendix:protocol-init}.

In Fig.~\ref{fig:scheme}~(c) we illustrate the Rabi protocol $\Omega(t)$ and the Raman protocols $A_j(t)$ of a high fidelity CNOT realization. In Fig.~\ref{fig:scheme}~(d) we show the transformation corresponding to this high fidelity realization. The CNOT transformation is visible in the computational subspace, while the transformation on the remaining subspace is arbitrary.

\section{Weak-Coupling Solutions}
\begin{figure*}
    \centering
    \includegraphics{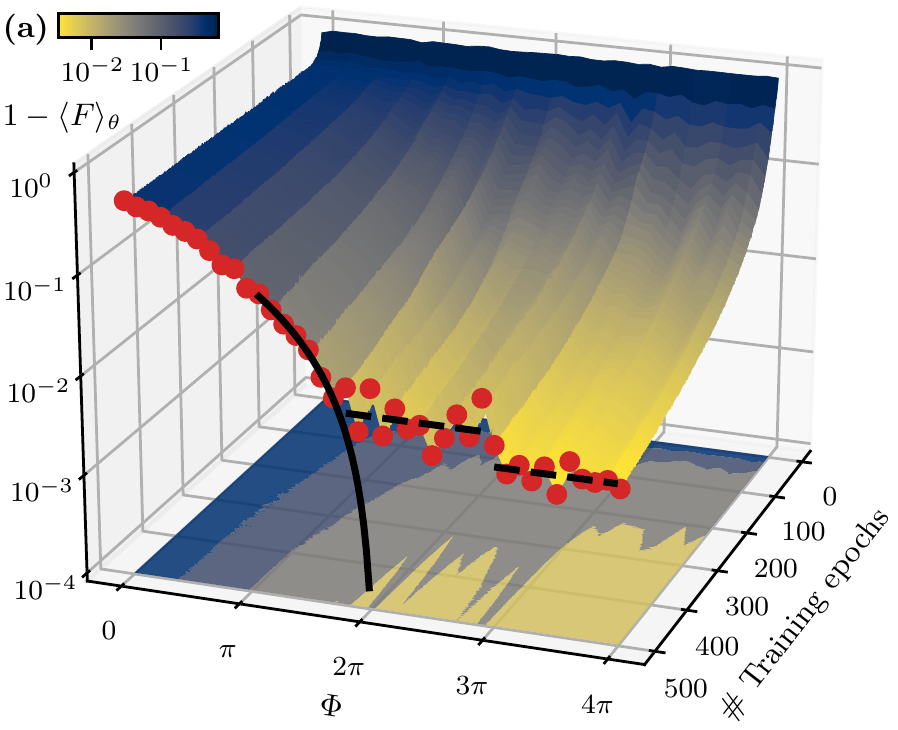}
    \includegraphics{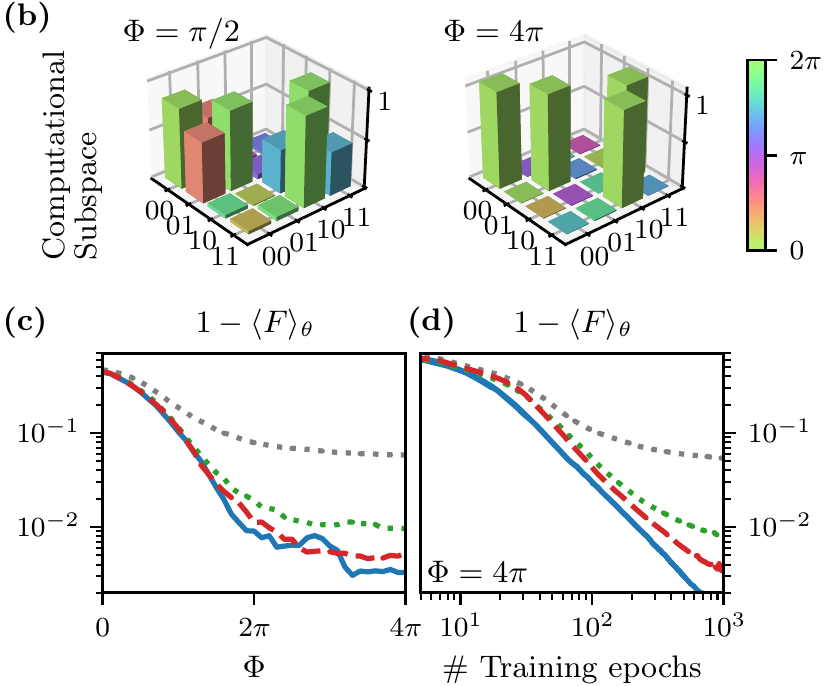}
    \caption{\textbf{Gate optimization in the weak-coupling limit.}
    \textbf{(a)} We display the infidelity $1-\langle F \rangle_{\theta}$ as a function of the gate action $\Phi$ and the number of training epochs. The infidelity after 500 training epochs is depicted as red dots. Near $\Phi_c\approx 2\pi$, we fit the infidelity $1-\langle F \rangle_{\theta}$ with the fitting function $\Phi=\sqrt{(1-\langle F \rangle_{\theta})/A}+\Phi_c$, which we depict as a black solid line, while the black-dashed lines depict the converged infidelities of $7\times 10^{-3}$ for $2\pi<\Phi<3\pi$ and $3\times 10^{-3}$ for $\Phi > 3\pi$.
    \textbf{(b)} Optimized gate transformations in the computational subspace. For a gate action of $\Phi=\pi/2$ the infidelity is $1-\langle F \rangle_{\theta}=0.25$, indicating that the gate action is insufficient to create a high fidelity protocol. For $\Phi=4\pi$ we show a transformation having an infidelity of $1-\langle F \rangle_{\theta}=3\times 10^{-4}$, indicating sufficient gate action. The phases of the matrix elements are encoded on a cyclic color map.
    \textbf{(c-d)} Infidelity without dissipation (blue) and with dissipation of values $1/\gamma=500\si{\micro\second}$ (red-dashed), $1/\gamma=100\si{\micro\second}$ (green-dotted) and $1/\gamma=10\si{\micro\second}$ (grey-dotted).
    \textbf{(c)} Infidelity after 500 training epochs as a function of the gate action $\Phi$.
    \textbf{(d)} The infidelity during training for a fixed gate action $\Phi=4\pi$ as a function of the number of training epochs. Dissipation results in slower reduction of the infidelity with the number of training epochs and determines the lower bound of the infidelity that is visible for large dissipation, i.e. for large $\gamma$.
    }
    \label{fig:training}
\end{figure*}
In this section, we identify optimal implementations of the CNOT gate in  the weak-coupling limit, i.e. based on dynamical phase accumulation. We consider a large interatomic distance $r\approx 10\si{\micro\metre}$, of the two atoms. For this distance, the van-der-Waals interaction is small compared to the maximal Rabi frequency $V_\text{vdW} \ll \hbar\Omega_\textrm{max}$ allowing for occupation of the Rydberg-Rydberg state $\ket{R}\otimes\ket{R}$, where $V_\text{vdW}=V_\text{vdW}^{1,2}(r)$, based on the van-der-Waals interaction in Eq.~\ref{eq:vdw}, with $r = |r_1-r_2|$.
Hence, the magnitude of the nonlinearity is limited by the algorithm time $\tau$ and the interaction strength $V_\text{vdW}$.
We introduce the dimensionless gate action
\begin{equation}
    \Phi=\tau V_\text{vdW}/\hbar,
\end{equation}
as the maximally achievable non-linear phase accumulation. Note that as the algorithm time $\tau=1\si{\micro\second}$ is fixed, the gate action $\Phi$ is equivalently a measure of the interaction strength.
In this section we treat the interaction strength $V_\mathrm{vdW}$ as an external parameter rather than a trainable parameter.

In Fig.~\ref{fig:training}~(a) we show the average of the infidelity $1-\langle F \rangle_\theta$ over 15 optimized protocols~\footnote{We choose the number of optimized protocols, with randomly sampled initial protocols, empirically.} for the target transformation of the CNOT gate. 
We show this as a function of the gate action $\Phi$ and the number of training epochs in the absence of dissipation, i.e. for $\gamma = 0$.
We find that for small gate actions $\Phi \lesssim 2\pi$ the optimization algorithm does not generate a high-fidelity protocol. The fidelity steadily increases with increasing gate action $\Phi$. We fit the expression $1-\langle F \rangle_{\theta}=A(\Phi-\Phi_c)^2$ in the vicinity of the critical gate action $\Phi_c$ and find that $A=0.37$ and $\Phi_c = 2.018\pi$.
For values of $2\pi<\Phi<3\pi$, the infidelity converges to approximately $1-\langle F \rangle_\theta \approx 7\times 10^{-3}$, which indicates sufficient gate action $\Phi$, i.e. it indicates that sufficient time and interaction is provided to generate a two-qubit operation. 
For gate actions of values $\Phi>3\pi$ the infidelity decreases further as it converges to approximately $1-\langle F \rangle_\theta\approx3\times 10^{-3}$. In this regime, we observe more efficient optimization behavior that reaches values of $1-\langle F\rangle_\theta < 10^{-2}$ after roughly $200$ training epochs.
In the case of no interaction, the transformation consists of single-qubit transformations that cannot implement the $\mathrm{CNOT}$ operation.

In Fig.~\ref{fig:training}~(b) we show representations of transformations in the computational subspace for $\Phi=\pi/2$ and $\Phi=4\pi$.
In the case of insufficient interaction strength, for $\Phi=\pi/2$, the implemented transformation is visibly distinct from a CNOT gate.
In the case of sufficiently large interaction strength, for $\Phi=4\pi$, a high fidelity implementation of the $\mathrm{CNOT}$ gate is visible.
In Fig.~\ref{fig:training}~(c) we show the infidelity of optimized protocols as a function of the gate action $\Phi$, with and without dissipation. We use the dissipative parameters discussed in Sect. II. For small values of $\Phi$, the gate fidelity is independent of dissipation, as the Rydberg state is weakly occupied during the protocol.
For increasing values of $\Phi$, the protocols approach high fidelities with dissipation, but with an increased infidelity. This increase of the infidelity is also visible along the learning trajectory for $\Phi=4\pi$ as we show in Fig.~\ref{fig:training}~(d).
Here we see that dissipation results in a lower bound of the infidelity of the optimized protocol.
This lower bound is reduced by minimizing the occupation time of the Rydberg states $\ket{R}_j$. The maximal Rabi frequency $\Omega_\textrm{max}$ provides a limitation of this optimization in the case of a fixed algorithm time $\tau$.

We find that the optimization method provides high fidelity protocols in the presence of experimentally motivated dissipation. Generally, higher fidelities than what we present can be achieved by increasing the number of training epochs. In a realistic setup, measurement noise, laser phase- and intensity noise, and spatial fluctuations are additional challenges, that can be included in our optimization approach.

\section{Spatial Fluctuations}
To demonstrate the robustness properties of the optimal implementations that we have obtained, we include fluctuations of the distance between the two Rydberg atoms. In an experimental realization, these fluctuations might derive from thermal motion of each of the atoms in the tweezer potentials, or fluctuations of the tweezer potential itself. We consider spatial distances between the two atoms of $4\si{\micro\meter}$ to $11\si{\micro\meter}$, which interpolates between the blockade regime and the weak-coupling limit.

We consider a high-fidelity implementation $U$ of the CNOT gate, which has been optimized for a specific interaction strength $V_\textrm{vdW}$ and zero dissipation $\gamma=0$.
We now introduce fluctuations of the atom distance, i.e. $r \rightarrow r + \delta r(t)$, in which the spatial fluctuations $\delta r(t)$ are sampled from a normal distribution $N(0, \sigma_r)$ at a frequency of $512\si{\mega\hertz}$.
Based on a single, stochastic time series $r + \delta r(t)$, we now determine the modified, time-dependent interaction strength
\begin{equation}
    V_\mathrm{vdW}(t) = \frac{\hbar C_6}{|r+\delta r(t)|^6}.
\end{equation}
We use this interaction strength to generate the time-evolution $\tilde{U}(\sigma_r)$, while keeping all other features of the protocol unchanged, i.e. we use the same $\Omega(t)$, $A_j(t)$ and $B$ protocol. To quantify to what degree the fidelity is reduced due to the spatial fluctuations, we define the average transformation error as 
\begin{equation}
    \epsilon(\sigma_r) = 1-\frac{1}{4}\langle|\text{Tr}(\tilde U(\sigma_r)^\dagger \mathrm{CNOT})|\rangle_{\delta r(t), \theta}.
\end{equation} 
Here we take the statistical average of the implementation error over $50$ sampled trajectories of $\delta r(t)$ and $10$ high fidelity protocols provided by transformation parameters $\theta$ optimized from different initial values. 

\begin{figure}
    \centering
    \includegraphics{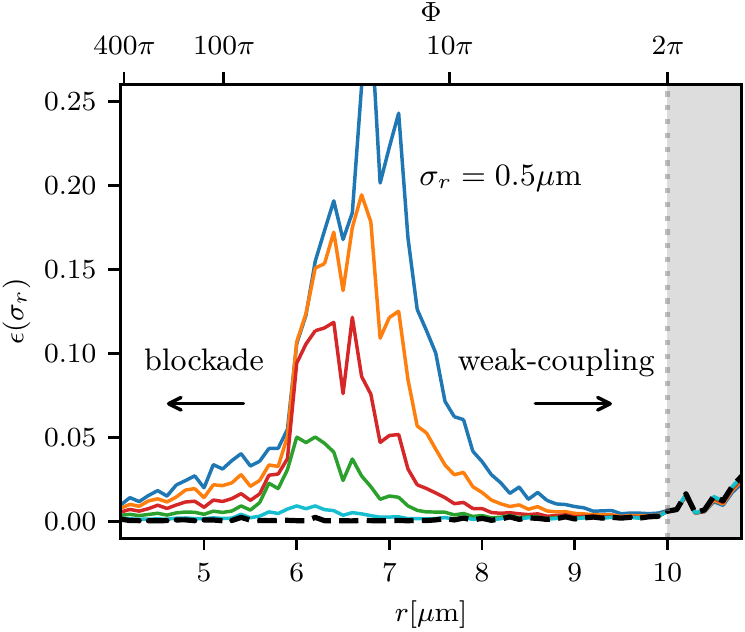}
    \caption{\textbf{Robustness against spatial fluctuations.}
    The transformation error $\epsilon(\sigma_r)$ in the presence of spatial fluctuations, for the noise parameter $\sigma_r = 0.5, 0.4, 0.3, 0.2, 0.1, 0\si{\micro\metre}$ (blue, orange, red, green, cyan, black dashed), and as a function of the interatomic distance $r$.
    For small ($<6\si{\micro\metre}$) and large ($>8\si{\micro\metre}$) distances the optimal implementations are robust against spatial fluctuations. For intermediate distances of $r=6-8\si{\micro\metre}$, the system is strongly susceptible to spatial fluctuations and the relative error grows quickly as a function of $\sigma_r$.
    The vertical dashed line depicts the minimum distance required to implement the $\mathrm{CNOT}$ gate, see Fig.~\ref{fig:training} for reference.}
    \label{fig:r-error}
\end{figure}
In Fig.~\ref{fig:r-error} we show the transformation error $\epsilon(\sigma_r)$ as a function of the interatomic distance $r$ for various values of the standard deviation $\sigma_r$. 
At large distances of about $r>9\si{\micro\metre}$, the protocols are only weakly susceptible to spatial fluctuations.
Since the van-der-Waals interaction in Eq.~\ref{eq:vdw} scales with $r^{-6}$, the gradient falls off rapidly as the mean distance $r$ increases. 
Because of this rapid fall-off, fluctuations of $r$ result in a smaller and smaller increase of the error $\epsilon(\sigma_r)$ with increasing $r$.
On the other hand, in the blockade regime at distances of $r < 5\si{\micro\metre}$, the interaction strength $V_\mathrm{vdW}$ dominates the maximal Rabi frequency $\Omega_\textrm{max}$. In this limit, transitions into the Rydberg-Rydberg state $\ket{R}_1\ket{R}_2$ are highly suppressed. Therefore, the spatial fluctuations do not induce large errors in this limit either but are more noticeable than in the weak-coupling limit, in this example.
However, in the intermediate regime of $5 \si{\micro\metre} < r < 8\si{\micro\metre}$, the transformation error $\epsilon(\sigma_r)$ is highly susceptible to spatial fluctuations. 
At these interatomic distances, the van-der-Waals interaction and the the maximal Rabi frequency are of the same order, i.e. $V_\textrm{vdW} \sim \hbar \Omega_\textrm{max}$.
Hence, the optimized protocols are highly susceptible to spatial fluctuations, making these intermediate interatomic distances undesirable in any realization.
The robustness with respect to spatial fluctuations is one of the key features that makes the Rydberg blockade regime attractive for quantum computing purposes~\cite{jaksch_fast_2000}.
However, we emphasize that in the weak-coupling limit, the system is equally robust against spatial fluctuations
\footnote{We note that if we expand this optimization to include the optimization of interatomic distances or spatial trajectories
of the tweezer locations, this effect may
be important to consider. The susceptibility to errors at
intermediate distances will create learning pressure away
from protocols that cross from the weak-coupling limit at
large distances to the Rydberg blockade regime at small
distances potentially inhibiting the convergence of the optimization.}.

\section{Conclusion}
In conclusion, we have demonstrated quantum gate optimization of a CNOT gate in a Rydberg architecture under experimentally motivated constraints, via machine learning.
The two qubit states are two long-lived hyperfine states of each of the two atoms. Additionally, we include a Rydberg state in each atom in our model, as an auxilliary state to provide a van-der-Waals interaction to create a two-qubit gate. These atoms are held in optical tweezers, at fixed distance. The model and parameter choices are based on 171-Yb, such as the dissipative properties of the Rydberg state. However, we emphasize that our approach is universally applicable to Rydberg architectures. We assume that the long-lived hyperfine states can be driven by Raman protocols, and the transition from one of the hyperfine states to the Rydberg state by a single global Rabi protocol. We show that utilizing either individual Raman protocols for each atom and a global Rabi protocol for both atoms, or individual Rabi protocols for each atom and a global Raman protocol for both atoms, is sufficient for universal quantum computing. Focusing on the case of individual Raman protocols for each atom and a global Rabi protocol, we utilize a hybrid quantum-classical optimization approach, based on gradient ascent pulse engineering (GRAPE), to determine protocols that implement a high fidelity CNOT gate. Keeping the total algorithm time of the protocols fixed at $1\si{\micro\second}$, we scan the optimal implementations as a function of the interaction strength. Finally, we map out the robustness of the optimal protocols against spatial fluctuations of the interatomic distance. We find that both for the weak-coupling limit and for the blockade regime, the implementations are robust. However, the intermediate regime, at which the maximal Rabi frequency is comparable to the van-der-Waals interaction, is not robust and thus undesirable. Additional imperfections, such as doppler shifts or imperfections in the laser intensities, will be explored elsewhere.
We also note that the weak-coupling regime enables gate implementations in tweezer arrays with strongly suppressed next-nearest interactions, resulting in more straight-forward implementations.
With these results, we have demonstrated the immediate and significant impact that hybrid quantum-classical optimization, or machine learning inspired methods in general, have on quantum gate design. Going forward, this suggests systematic, large-scale, and in-depth utilization of quantum machine learning methods.

\begin{acknowledgments}
This work is funded by the Deutsche Forschungsgemeinschaft (DFG, German Research Foundation) - SFB-925 - project 170620586 and the Cluster of Excellence 'Advanced Imaging of Matter' (EXC 2056) project 390715994.
\end{acknowledgments} 

\bibliography{main.bib}

\appendix
\section{Protocols and parameterization}
\label{appendix:protocol-parameterization}
In the following we detail the parameterization of the protocols in our optimization method. 
We denote the concatenated parameters of the protocol as $\theta=\{\theta_\Omega, \theta_{\partial\phi}, \theta_{A_1}, \theta_{A_2}, \theta_B\}$.
We represent the protocols $|\Omega(t)|$ and $A_j(t)$ in a step-wise discretized manner such that the elements of $\vartheta\in \theta_\Omega, \theta_{A_j}$ represent the amplitudes of respective protocols at $m$ discrete points in time. 
We linearly interpolate these step-wise representations $\vartheta\in\mathbb{R}^m$ on the temporal lattice with the step-size $\Delta t=\tau/(m-1)$. The interpolation is
\begin{equation}
    s(\vartheta, t) = (1-p_i)\vartheta_i + p_i\vartheta_{i+1},
\end{equation}
where $i=\lfloor t/\Delta t \rfloor$ is the latest index corresponding to the time $t$ and $p_i=t/\Delta t -i$ is an interpolation weight. The amplitudes of the Rabi protocol $|\Omega(t)|$, the control Raman protocol $A_1(t)$ and the target Raman protocol $A_2(t)$ are then given by
\begin{align}
    |\Omega(t)| &= s(\theta_\Omega, t) \\
    A_1(t) &= s(\theta_{A_1}, t) \\
    A_2(t) &= s(\theta_{A_2}, t).
\end{align}
The phase $\phi(t)$ of the Rabi protocol $\Omega(t)$ is given by the stepwise differential parameterization
\begin{equation}
    \phi(t) = s(\theta_{\phi}, t),
\end{equation}
where $\theta_{\phi, i} = \theta_{\phi, i-1}+\theta_{\partial\phi,i}$ and $\theta_{\phi,0}=\theta_{\partial\phi,0}$. 
This construction creates slowly varying phase protocols and avoids sudden phase-jumps.
The magnetic field is given by the constant parameterization
\begin{equation}
    B = \theta_B.
\end{equation}
In the presented analysis we consider a total number of $4\times 64+1$ parameters, that is $\theta\in\mathbb{R}^{257}$.
Further, as mentioned in the main text, we constrain the parameters in between minimal and maximal values.
The constraints $g_i(\theta_i)$ in Eq.~\ref{eq:param-step} are defined as follows
\begin{align}
    g_{\Omega}(\theta_{\Omega,i}) &= \max(0, \min(\Omega_\textrm{max}, \theta_{\Omega,i})) \\
    g_{A_1}(\theta_{A_1,i}) &= \max(0, \min(A_\textrm{max}, \theta_{A_1,i})) \\
    g_{A_2}(\theta_{A_2,i}) &= \max(0, \min(A_\textrm{max}, \theta_{A_2,i})) \\
    g_{\partial\phi}(\theta_{\partial_t\phi,i}) &= \max(\partial_t\phi_\textrm{min}, \min(\partial_t \phi_\textrm{max}, \theta_{\partial\phi,i})) \\
    g_{B}(\theta_{B}) &= \max(B_\textrm{min}, \min(B_\textrm{max}, \theta_{B})),
\end{align}
with $\Omega_\textrm{max} = 2\pi\times 10 \si{\mega\hertz}$, $A_\textrm{max}=2\pi\times 10 \si{\mega\hertz}$, $\partial_t\phi_\mathrm{min}=-\pi/100\si{\nano\second}$, $\partial\phi_\mathrm{min}=+\pi/100\si{\nano\second}$, $B_\mathrm{min}=100\mathrm{G}$ and $B_\mathrm{max}=200\mathrm{G}$.

\section{Protocol initialization}
\label{appendix:protocol-init}
\begin{figure}
    \centering
    \includegraphics{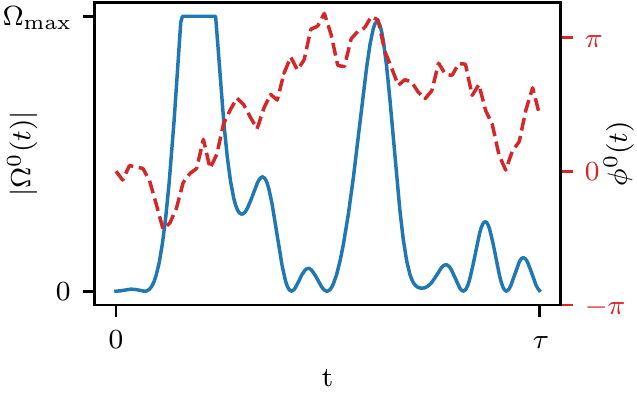}
    \caption{\textbf{Protocol initialization.} The initial Rabi protocol $\Omega(t)$ of algorithm time $\tau=1\si{\micro\second}$. The amplitude of a sample initial Rabi protocol $|\Omega^0(t)|$ (blue) contains 16 sinosiodal modes and has a maximal Rabi frequency of $\Omega_\textrm{max}$. The initial Raman protocols $A_1^0(t)$ and $A_2^0(t)$ are initialized by the same strategy. The phase of the initial Rabi protocol $\phi^0(t)$ (red-dashed) is given by a random walk starting at $\phi^0(0)=0$.}
    \label{fig:protocol-init}
\end{figure}
We construct the initial parameters such that the resulting protocols vary slowly, start and end at zero, and are fairly well-behaved. We first consider a distribution of initial parameterizations
\begin{equation}
    S = \left\{ \sum_k \phi_k \sin(k\pi t/\tau)\right\},
\end{equation}
where $\phi_k$ are random numbers from the uniform distributions $[-1/\sqrt{k}, +1/\sqrt{k}]$. We introduce this dependence on $k$ to emphasize slow modes.
We initialize the transformation parameters $\theta_\Omega$, $\theta_{A_1}$ and $\theta_{A_2}$ such that the following initial protocols are realized
\begin{align}
    |\Omega^0(t)| &= \Omega_\text{init}  \min(1, s^2_\Omega(t)) \\
    A_1^0(t) &= A_\text{init}\min(1, s^2_{A_1}(t)) \\
    A_2^0(t) &= A_\text{init}\min(1, s^2_{A_2}(t)),
\end{align}
where we sample $s_\Omega, s_{A_1}$ and $s_{A_2}$ from the distribution $S$, $\Omega_\text{init}$ is the maximal frequency of the initial Rabi protocol and $A_\text{init}$ is the maximal frequency of the initial Raman protocols. 
Through out this work we scale the initial protocols by the corresponding maximal frequencies, i.e. $\Omega_\text{init}=\Omega_\textrm{max}$ and $A_\text{init}=A_\textrm{max}$.
The parameters of the phase are initialized as $\theta_{\partial_\phi, 0}^0 = 0$ and $\theta_{\partial_\phi, i}^0$ is uniformly sampled from $[-\delta, \delta]$.
where $\delta\in\mathbb{R}^+$. Here we choose $\delta = 1.5$.
Fig.~\ref{fig:protocol-init} shows an example of a random initial Rabi protocol $\Omega^0(t)$.
The initial magnetic field $B^0$ is sampled from the distribution $[B_\textrm{min}, B_\textrm{max}]$.

\section{Maximal Rabi frequency}
\label{appendix:omega-max}
\begin{figure}
    \centering
    \includegraphics{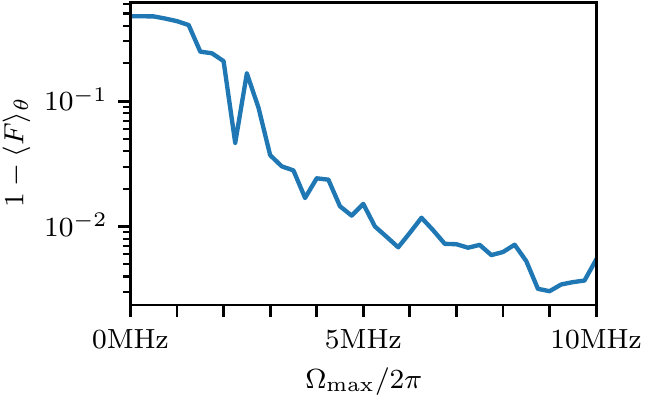}
    \caption{\textbf{Maximal Rabi frequency.} High fidelity protocols are realized for $\Omega_\textrm{max} = 2\pi \times 10\si{\mega\hertz}$, with a gate action of $\Phi=4\pi$ after 400 training epochs. Lower infidelities can be achieved with more training epochs.}
    \label{fig:pulse-infidelity}
\end{figure}

The maximal Rabi frequency $\Omega_\textrm{max}$ provides a limitation on the minimally achievable infidelity $1-F$ of the gate protocol. If $\Omega_\textrm{max}$ is not sufficiently large to complete a Rabi oscillation of one of the states $\ket{1}_1$ or $\ket{1}_2$ to one of the Rydberg states $\ket{R}_1$ or $\ket{R}_2$ during the protocol of algorithm time $\tau$, then no protocols with satisfactory fidelity can be constructed.
In Fig.~\ref{fig:pulse-infidelity} we show the average infidelity $1-\langle F \rangle_\theta$ over several optimized protocols as a function of the Rabi frequency $\Omega_\textrm{max}$ for a fixed maximal Raman frequency of $A_\textrm{max}=2\pi \times 10\si{\mega\hertz}$ and a gate action of $\Phi=4\pi$.
For small values of the maximal Rabi frequency $\Omega_\textrm{max} < 2\pi \times 2 \si{\mega\hertz}$, the infidelity displays a plateau of large values of roughly $1-\langle F \rangle_\theta\approx 0.5$.
With increasing $\Omega_\textrm{max} > 2\pi \times 2\si{\mega\hertz}$, the infidelity decreases.
For $\Omega_\textrm{max} > 2\pi \times 8\si{\mega\hertz}$, the infidelity begins to saturate at values $1-\langle F \rangle_\theta < 10^{-2}$. Note that the protocols are not necessarily fully converged and lower infidelities can be achieved with more training epochs. Therefore, a maximal Rabi frequency $\Omega_\textrm{max} = 2\pi \times 10\si{\mega\hertz}$ is sufficient for our analysis.

\section{Universal Quantum Computing with Global Pulses}
\label{appendixa}
In the following we demonstrate that a single global Rabi coupling for $N_a$ neutral atoms is capable of universal quantum computing. We also show this for the case of a single global Raman coupling and invidiual Rabi couplings.
We consider $N_a$ three-level systems consisting of the states $\{\ket{R}, \ket{1},\ket{0}\}$, and the Hamiltonian
\begin{equation}
H = \mu B \sum_{j=1}^n \sigma^j_z + \sum_{j=1}^n A_j(t)\sigma^j_x + \sum_{j=1}^n (\Omega^j_x(t) \tau^j_x+\Omega^j_y(t) \tau^j_y) + H_I,
\label{Ham}
\end{equation}
where 
\begin{align}
    \sigma^j_x = \begin{pmatrix}
        0 & 0 & 0 \\
        0 & 0 & 1 \\
        0 & 1 & 0 \\
    \end{pmatrix}&&
    \sigma^j_y = \begin{pmatrix}
        0 & 0 & 0 \\
        0 & 0 & -i \\
        0 & i & 0 \\
    \end{pmatrix}&&
    \sigma^j_z = \begin{pmatrix}
        0 & 0 & 0 \\
        0 & 1 & 0 \\
        0 & 0 & -1 \\
    \end{pmatrix},
\end{align}
act on the subspace $\{\ket{1},\ket{0}\}$ of the $j$th system.
\begin{align}
    \tau^j_x = \begin{pmatrix}
        0 & 1 & 0 \\
        1 & 0 & 0 \\
        0 & 0 & 0 \\
    \end{pmatrix}&&
    \tau^j_y = \begin{pmatrix}
        0 & -i& 0 \\
        i & 0 & 0\\
        0 & 0 & 0 \\
    \end{pmatrix}&&
    \tau^j_z = \begin{pmatrix}
        1 & 0 & 0 \\
        0 & -1& 0 \\
        0 & 0 & 0 \\
    \end{pmatrix},
\end{align}
act on the subspace $\{\ket{R},\ket{1}\}$ of the $j$th atom.
Analogously, we also define the matrices on the subspace $\{\ket{R},\ket{0}\}$ of the $j$th atom as
\begin{align}
    \nu^j_x = \begin{pmatrix}
        0 & 0 & 1 \\
        0 & 0 & 0 \\
        1 & 0 & 0 \\
    \end{pmatrix}&&
    \nu^j_y = \begin{pmatrix}
        0 & 0 & -i \\
        0 & 0 & 0 \\
        i & 0 & 0 \\
    \end{pmatrix}.
\end{align}
The interaction term is 
\begin{equation}
    H_I = \sum_{\langle i,j\rangle } \frac{C_6}{(r_i(t)-r_j(t))^6} \ket{R}_i \otimes \ket{R}_j \bra{R}_i \otimes \bra{R}_j,
\end{equation} 
where $r_j(t)$ is the real-space position of the $j$th atom.
$\mu B$ is the Zeeman splitting due to a constant and global magnetic field $B$. 
$A_j(t)$ is the amplitude of the $j$th Raman coupling of the $j$th atom that control the transition between $\ket{1}_j$ and $\ket{0}_j$.
$\Omega^j_{x,y}(t)$ are the Rabi coupling components of the $j$th atom that control the transition between $\ket{1}_j$ and $\ket{R}_j$.
For convenience we denote the global sums of local operators as $S_z=\sum_{j=1}^n\sigma_z^j$, $T_x = \sum_{j=1}^n \tau^j_x$, $T_y = \sum_{j=1}^n \tau^j_y$ and $V=\sum_{i,j}V_{i,j}$.
We consider the base set of generators contained in Eq.~\ref{Ham},
\begin{align}
   \mathcal{H}'_0 = \{\sigma_x^1,\dots,\sigma_x^n, S_z, \tau_x^1, \dots, \tau_x^n,\tau_y^1, \dots, \tau_y^n, V \}.
\end{align}
We consider that $B>0$ is always on, which makes controlling individual local rotations more difficult.
We consider the rotating frame given by $U=\exp\{i \frac{1}{2}\mu B t S_z\}$, such that we reduce the base set of generators to 
\begin{equation} 
    \mathcal{H}''_0 = \{\sigma_x^1,\dots,\sigma_x^n, \tau_x^1, \dots, \tau_x^n,\tau_y^1, \dots, \tau_y^n, V \},
\end{equation}
with all generators now being controllable individually from each other.

First, we consider the case in which $\Omega^j_{x,y}=\Omega_{x,y}$, such that the local operators $\tau_{x,y}^i$ are no longer individually controllable. 
The base set of generators becomes 
\begin{equation} 
    \mathcal{H}_0 = \{\sigma_x^1,\dots,\sigma_x^n, T_x, T_y, V \}\label{H0}.
\end{equation}
From these base generators we find the commutators
\begin{align}
[\sigma_x^i,T_x]&=\sum_{j=1}^n[\sigma_x^i, \tau_x^j]=-i \nu_y^i\label{nuy}\\
[\nu_y^i,\sigma_x^i]&=-i\tau_x^i\label{taux}\\
[\nu_y^i,T_y]&=\sum_{j=1}^n[\nu_y^i, \tau_y^j]=-i \sigma_y^i.\label{sigy}
\end{align}

Eqs.~\ref{nuy} and \ref{taux} can be repeated analogously to obtain $\nu_x^i$ and $\tau_y^i$. 
This means that despite the global Rabi term that determines the transition between $\ket{1}_j$ and $\ket{R}_j$ for all $1 \leq j \leq N_a$, the local generators $\tau_{x,y}^i$ are part of the dynamical Lie algebra and therefore controllable individually.
From Eq.~\ref{sigy} we see that that $\sigma_y^i$ is accessible, and therefore $\sigma_z^i$ is accessible as well. This allows full access to local single-qubit operations.
Note that this means the magnetic field $B$ was not necessary for computational purposes to begin with. 
However, in experimental setups it serves the purpose of providing non-degenerate levels $\ket{0}$ and $\ket{1}$ at all times.
Since the transformation $U$ into the comoving frame  does not affect the interaction term, i.e. $UVU^\dagger=V$, we find that the base set of generators in Eq.~\ref{H0} is computationally universal on the logical space $\otimes_{j=1}^n \{\ket{0}^j,\ket{1}^j\}$.
We demonstrate this by constructing specific examples of rotations around $\tau_x^i$, $\tau_y$, $\sigma_y^i$ and $\sigma_z^i$ by an arbitrary angle $\alpha$ as
\begin{align}
e^{i\alpha\tau_x^i} &= e^{i\frac{3\pi}{2} \sigma_x^i}e^{i \frac{3\pi}{2}T_x}e^{i\alpha \sigma_x^i}e^{i\frac{\pi}{2}T_x}e^{i\frac{\pi}{2} \sigma_x^i}\\
e^{i\alpha\tau_y^i} &= e^{i\frac{3\pi}{2} \sigma_x^i}e^{i \frac{3\pi}{2}T_y}e^{i\alpha \sigma_x^i}e^{i\frac{\pi}{2}T_y}e^{i\frac{\pi}{2} \sigma_x^i}\\
e^{i \alpha\sigma_y^i} &= e^{i\frac{3\pi}{2} \sigma_x^i}e^{i\frac{3\pi}{2}T_y}e^{i \frac{3\pi}{2}T_x}e^{i\alpha \sigma_x^i}e^{i\frac{\pi}{2}T_x}e^{i\frac{\pi}{2}T_y}e^{i\frac{\pi}{2} \sigma_x^i}\\
e^{i \alpha\sigma_z^i} &= e^{i\frac{7\pi}{4} \sigma_x^i}e^{i\frac{3\pi}{2}T_y}e^{i \frac{3\pi}{2}T_x}e^{i\alpha \sigma_x^i}e^{i\frac{\pi}{2}T_x}e^{i\frac{\pi}{2}T_y}e^{i\frac{\pi}{4} \sigma_x^i}.
\end{align}
From these rotations, entanglement between qubits can be achieved utilizing $V$ in the canonical manner of Rydberg architectures.

Second, we consider the case of individual Rabi couplings $\Omega^j_{x,y}(t)$, but a single global Raman coupling $A_j(t)=A(t)$.
Analogously to the previous case, the base set of generators then becomes 
\begin{equation}
\mathcal{H}_0 = \{S_x, \tau_x^1,\dots,\tau_x^n,\tau_y^1,\dots,\tau_y^n,V\}\label{h0raman}.
\end{equation}
The argument follows analogously and we find that 
\begin{align}
[S_x,\tau_x^j]&=\sum_{i=1}^n[\sigma_x^i, \tau_x^j]=-i \nu_y^j\label{nuy2}\\
[\nu_y^i,\tau_x^i]&=i\sigma_x^i\label{taux2}\\
[\nu_y^i,\tau_y^i]&=-i \sigma_y^i\label{sigy2},
\end{align}
and therefore arbitrary single-qubit rotations can be constructed as
\begin{align}
e^{i \alpha\sigma_x^i} &= e^{i \frac{3\pi}{2} \tau^i_x}e^{i \frac{3\pi}{2} S_x}e^{i \alpha \tau^i_x}e^{i \frac{\pi}{2} S_x}e^{i \frac{\pi}{2} \tau^i_x}\\
e^{i \alpha\sigma_y^i} &= e^{i \frac{3\pi}{2} \tau^i_x}e^{i \frac{3\pi}{2} S_x}e^{i \alpha \tau^i_y}e^{i \frac{\pi}{2} S_x}e^{i \frac{\pi}{2} \tau^i_x}\\
e^{i \alpha\sigma_z^i} &= e^{i \frac{3\pi}{2} \tau^i_x}e^{i \frac{3\pi}{2} S_x}e^{i \frac{\pi}{4} \tau^i_y}e^{i \alpha \tau^i_x}e^{i \frac{7\pi}{4} \tau^i_y}e^{i \frac{\pi}{2} S_x}e^{i \frac{\pi}{2} \tau^i_x},
\end{align}
despite only global control over $S_x$. 
It again follows that the dynamical Lie algebra is capable of all necessary operations for universal quantum computing.
Note in particular that the generator $S_z$ associated with the magnetic field was again not necessary for constructing arbitrary single-qubit rotations.

In this case of global Raman coupling, the Rabi couplings can be performed individually which means that there is no undesirable population in Rydberg states as overhead of unrelated transformations.
The overhead transformation occurs only on the local $\sigma_x$ which is easily circumvented.
We can perform a CNOT gate on the $1$st and $2$nd qubit while in the Rydberg blockade radius as 
\begin{align}
\mathrm{CNOT}&=
e^{i\frac{\pi}{2}\tau_x^2}
R
e^{i \pi \tau_x^1}
R
e^{i\frac{\pi}{2}\tau_x^2}\label{app:CNOT1}\\
R&=
e^{i\frac{\pi}{2}S_x}
e^{i\frac{\pi}{4}\tau^2_x}
e^{i\frac{\pi}{\sqrt{8}}(\tau^2_x+\tau_y^2)}
e^{i\frac{\pi}{4}\tau^2_x}
e^{i\frac{3\pi}{2}S_x},\label{app:CNOT2}
\end{align}
in the presence of any amount of other qubits which by construction also experience the rotations generated by $S_x$. This transformation will in particular act on all these additional qubits as the identity, as desired.
Finally, we note that the controlled-$Y$ gate is implemented more naturally in this case as
\begin{align}
\mathrm{C(Y)}&=e^{-i \frac{\pi}{2} \tau_x^2}e^{-i \frac{\pi}{4} S_x}e^{i\pi \tau_x^1}e^{i \frac{\pi}{4} S_x}e^{i \frac{\pi}{2} \tau_x^2}\label{app:CY1}.
\end{align}
Here it is implied that every transformation also contains the presence of the interaction term $V$ which is large enough to consider the system to be in the Rydberg blockade regime.
We emphasize that the Rydberg blockade is not necessary for the arguments we make about universal quantum computing. We only consider the Rydberg blockade for the analytically constructed examples in Eqs.~\ref{app:CNOT1}, \ref{app:CNOT2} and \ref{app:CY1}.

\end{document}